# An Analytics-based Decision Support System for Resource Planning under Heterogeneous Service Demand of Nursing Home Residents

Xuxue Sun, Nazmus Sakib, Nan Kong, Hongdao Meng, Kathryn Hyer, Chris Masterson, and Mingyang Li

## Abstract

Nursing homes (NHs) are critical healthcare infrastructures for caring frail older adults with 24/7 formal care and personal assistance. Adequate NH resource planning is of great importance to ensure desired quality of care and resident outcomes yet challenging. The challenge lies in the heterogeneous service demand of NH residents, due to the varied individual characteristics, the diverse dwelling duration with multiple competing discharge dispositions, and the diverse service need. Existing healthcare staffing literatures often assumed a homogeneous population of NH residents and neglected the complexity of service demand heterogeneity. This work proposes an analytics-based modeling framework with a user-friendly decision support platform for NH resource planning. The proposed framework characterizes the heterogeneous service demand of NH residents via novel integration of advanced statistical modeling, computer simulation and optimization techniques. We further provide a case study using real data from our industrial collaborator to demonstrate the effectiveness and superior performance of the proposed work. The impacts of service utilization heterogeneity and service need heterogeneity on resource planning decisions are investigated as well.

## Note to Practitioners

This work is motivated by the challenging task of nursing home resource planning under the complex and heterogeneous service demand of nursing home residents. This research work integrates predictive statistical models with computer simulation and optimization techniques to characterize heterogeneous service demand of NH residents at higher modeling fidelity and to facilitate resource planning decision making in response to service demand heterogeneity. The strategies and policies derived from the proposed decision support system are implementable and actionable in current nursing home industry practice. The proposed framework can easily adapt to varied residents census compositions across different nursing facilities or different seasonal periods with minimum adjustment. Moreover, the proposed methodological framework can be extended for resource preparedness and care delivery decisions under extreme event scenarios, such as influenza seasons or pandemic.

# I. INTRODUCTION

*A. Motivation*

The elderly population in the United States has a rapid growth in recent years. Along with the demographic shift in the elderly population, there is a considerable increase in the prevalence of aging-related disabilities, multiple chronic diseases and functional limitations/losses among older adults. Consequently, their need for nursing care and personal assistance is dramatically increasing in healthcare systems, especially in modern nursing homes (NHs) which are mainly responsible for caring frail and vulnerable elderly population. Modern NHs provide 24/7 formal care to frail older adults with a variety of skilled personal medical care and living assistance. Different types of nursing staff in NH, such as certified nurse assistant (CNA), registered nurse (RN), and license practical nurse (LPN), provide various care services, including post-acute health care, specialized medical services, restorative and rehabilitation services, assistance with activities of daily living and end- of-life care. Although there are multiple different types of nursing care supplies, the aging-related issues of NH residents consequently trigger escalating nationwide NH workforce shortage and insufficient care resources as well as public financing shortfall [1]. On one hand, the shortage in available beds and nursing staff will arouse concerns on care accessibility of NH residents [2], [3] and undermine their quality of life [4], [5]. Specifically, there are over 45 million Medicare-eligible aging adults with potential needs of NH care services, as compared to the insufficient nationwide supply of 1.7 million available NH beds [6]. Moreover, there are nationwide severe CNAs shortage with 90% understaffing. On the other hand, excessive supplies in capacity and workforce may lead to a waste of resource and increase financial burden of NH providers. The unused beds will incur unnecessary healthcare expenditures and overstaffing will reduce the profitability of NH providers. To meet with the excessive service demand of NH residents, it will be highly desirable to develop a cost-effective solution for NH resource planning which can ensure both the quality of care for NH residents and the profitability for NH providers.

Nevertheless, successful NH resource planning is challenging due to the highly heterogeneous and complex service demand of NH residents. The complexities of heterogeneous service demand lie in three aspects. First, different NH residents may have varied individual characteristics [7], such as diverse socio-demographics, various clinical diagnoses and chronic conditions, and different functional performance limitations (e.g., physical, mental and cognitive performance limitations and losses). Second, attributed to such diverse individual characteristics, the resident-level service utilization, such as dwelling duration, is heterogeneous among NH residents. Specifically, different NH residents may have diverse length-of-stay (LOS) [8] with multiple competing and correlated discharge dispositions (e.g., hospital, community, etc.) due to different individual characteristics. For example, NH residents may be discharged to community in a few weeks due to recovery, or stay in NH longer for several months or years due to further deterioration of their health conditions. Even during their stays, some NH residents may be transferred to acute care settings (e.g., hospital) due to the occurrence of critical events and emergencies (e.g., falls, infections, etc.). Third, the diverse health conditions and residents characteristics also lead to diverse service need among NH residents. Per diem staff-time of certain care service may vary among different NH residents. For example, daily CNA staff-time ranges from 30 to 200 minutes [9] among different individuals. Moreover, NH residents have diverse needs (e.g., skilled nursing care, restorative care, physical therapy, etc.) on different types of nursing staff and aides due to varied individual health conditions. The daily staff-time of different types of care services also varies among NH residents. In addition, the service need of NH residents may also changes over time due to the varying resident-level functional performance limitations and therapy intensity. There is a research need to improve NH capacity and workforce planning in response to such complex and highly heterogeneous service demand of NH residents.

*B. Related Works*

Existing NH resource planning approaches in nursing literature and industrial practice were mainly based on personal experiences and subjective judgment of NH administrators [10], [11], and/or often relied on government regulations, such as minimum staff-to-resident (SR) ratio enforced by Federal/state agencies [12], [13]. Many of them adopted one-size-fits-all policies based on aggregated measures, such as SR ratio and hours per patient day, which neglected service demand heterogeneity of NH residents. There was a lack of both data-driven framework and practical analytics-based decision support system to inform NH administrators of proper and managerial resource planning decisions. There is a need to develop an evidence-based analytical framework for NH capacity and workforce planning via taking the aforementioned complexities of service demand heterogeneity into account. On the other side, extensive works in the field of healthcare system and operations research focused on acute care settings, such as intensive care units [14], emergency units [15] and hospitals [16]-[22]. There is few research focusing on long-term care (LTC) settings [23], [24], such as LTC network and NH facilities. The existing studies often simplified the service demand via assuming a homogeneous population of NH residents and neglecting the complexity of heterogeneous service demand of NH residents [19], [20]. Many of them considered the distribution-based LOS model without incorporating individual characteristics [17]-[20], [23]. Some works utilized patient volume to quantify service demand without accounting for individual differences among patients [18], [20]. Further, most of existing methods failed to account for temporal dynamics and temporal heterogeneity in service demand modeling. There is a need to better understand and characterize the complex and heterogeneous service demand of NH residents, and further to develop an integrated evidence-based decision support system for NH capacity and workforce planning under heterogeneous service demand of NH residents.

To address the above research gaps and to meet with the research needs, we propose a data-driven integrated framework and analytics-based decision support system for NH resource planning in this chapter. The details will be described in the following sections. The remainder of the paper is organized as follows. Section. II describes the research problem and gives the overview of the proposed modeling framework. Section. III elaborates the developed predictive analytics integrated simulation approach to characterize the heterogeneous service demand of NH residents, followed by the demand-based two- phase optimization module for NH resource planning. Section. IV presents a real case study to illustrate the effectiveness of proposed modeling framework and to demonstrate its superior prediction performance as well as decision performance via comprehensive comparison. Section. V draws the conclusion of the proposed work.

## II.  METHODOLOGICAL FRAMEWORK

*A. Problem Description*

The goal of NH resource planning is to determine the proper amount of resources (e.g., number of beds and number of nursing staff) with desired criterion. Specifically, the goal of capacity planning is to determine the minimum required number of beds to meet with residents demand at the desired service quality level. Further, the goal of workforce planning is to determine the number of nursing staff need to be recruited to meet with facility-level service demand at reduced cost. However, due to the complex and heterogeneous service demand of NH residents, successful NH resource planning becomes challenging. This work aims to resolve the complexities of service demand of NH residents for adequate resource planning decisions.

*B. System Overview*

To determine the right amount of resources with proper scheduling in presence of heterogeneous service demand of NH residents, we propose a data-driven integrated modeling framework and an analytics-based decision support system. The proposed framework integrates advanced statistical model, computer simulation and stochastic optimization, as illustrated in Figure. 1. First, we develop a predictive individual NH LOS model to quantify heterogeneous service utilization of NH residents. Then, we develop a service need classification system with incorporated domain knowledge to characterize individual heterogeneity of service need of NH residents. With the developed predictive analytics, we further integrate with computer simulation to obtain facility-level service demand at higher granularity. To meet with the characterized service demand, we develop a two-phase demand-based optimization approach to achieve optimal capacity decision and optimal workforce planning. The details of proposed modeling framework and decision support system are elaborated in the following sections.

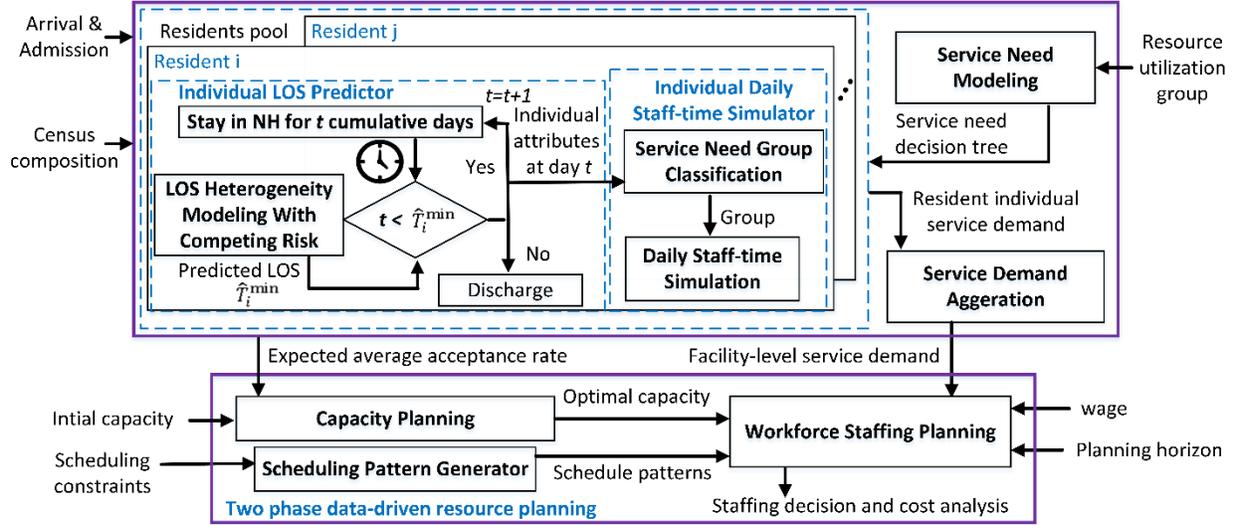

Fig. 1: Overview of the proposed modeling framework and decision support system

## III. ANALYTICS-BASED APPROACH FOR NURSING HOME RESOURCE PLANNING

*A. Length-of-Stay Heterogeneity Modeling With Competing Risk*

LOS is a critical measure of service utilization of NH residents. Due to the varied individual characteristics, LOSs differ considerably among residents. Further, NH residents will either be discharged back to their residential community due to recovery, or be transferred to higher-level acute care settings, such as hospital, due to the occurrence of critical events (e.g., fall, and infection, etc.). To improve the LOS prediction accuracy in presence of such individual heterogeneity with multiple competing discharge dispositions, we develop a semi-parametric predictive model with competing risk. The predictive model can better quantify and predict the service utilization of NH residents. In addition to the superior prediction accuracy, the developed model also achieves high model interpretability by identifying the disposition-specific contributing factors and quantifying their effects on NH service utilization. We will present the model formulation and estimation procedure as follows.

*1) Model Formulation:* Considering a heterogeneous population of *N* elderly NH residents and *M* discharge dispositions, we denote $T_{im}$ as LOS of resident *i* with discharge disposition *m*, $\forall i = 1...N, m = 1...M$. The proposed LOS model can be formulated as

$$\gamma_{im}(t|\boldsymbol{x}_i) = \frac{\lim_{\Delta t \to \infty} \Pr(t \leq T_{im} < t + \Delta t | T_i^{\min} \geq t, \boldsymbol{x}_i)}{\Delta t} = \gamma_m^b(t)\exp(\boldsymbol{\beta}_m^{\mathrm{T}}\boldsymbol{x}_i), \quad i = 1...N, m = 1...M \qquad (1)$$

where $T_i^{min} = \min(T_{i1}, \ldots T_{iM})$. $\gamma_{im}(\cdot)$ refers to the resident-specific NH discharge risk of disposition $m$ and $\gamma_m^b(\cdot)$ is the baseline NH discharge risk of disposition $m$, $m = 1\ldots M$. $\boldsymbol{\beta}_m$ and $\boldsymbol{x}_i$ represent the disposition-specific covariates coefficients and the individual characteristics (e.g., demographics, clinical diagnoses, functional performance, etc.), respectively. Based on the above formulation in Eq. (1), we can further estimate model parameters and quantify the effects of influencing factors.

*2) Model Estimation:* The original likelihood function has complex structure due to multiple discharge dispositions, making model parameters estimation challenging. To address this estimation issue, we introduce augmented variables $\Delta_{im}$, $\forall i = 1\ldots N$, $m = 1\ldots M$ to indicate the type of discharge disposition and further to decouple the likelihood function. $\Delta_{im} = 1$ if resident $i$ is discharged to disposition $m$ and $\Delta_{im} = 0$ otherwise. Given the augmented data $\boldsymbol{D} = \{t_i, \Delta_{im}, \boldsymbol{x}_i, \forall i = 1 \ldots N, m = 1 \ldots M\}$ and a set of unknown parameters $\boldsymbol{\Theta} = \bigcup_m \boldsymbol{\Theta}_m$ where $\boldsymbol{\Theta}_m$'s are mutually exclusive and $\boldsymbol{\Theta}_m = \{\gamma_m^b(\cdot), \boldsymbol{\beta}_m\}$, the original likelihood can be decomposed into multiple disposition-specific likelihood functions as

$$L(\boldsymbol{\Theta}|\mathbf{D}) = \prod_{i=1}^{N}\prod_{m=1}^{M} \left\{\gamma_m^b(t_i)\exp(\boldsymbol{\beta}_m^T \boldsymbol{x}_i)\exp\left[-\sum_{m=1}^{M}\int_0^{t_i}\gamma_m^b(\tau)\exp(\boldsymbol{\beta}_m^T \boldsymbol{x}_i)d\tau\right]\right\}^{\Delta_{im}}$$
$$= \prod_{m=1}^{M} L_m(\boldsymbol{\Theta}_m \mid \mathbf{D}) \qquad (2)$$

where $L_m(\boldsymbol{\Theta}_m \mid \boldsymbol{D}) = \prod_{i \in I_m}\gamma_m^b(t_i)\exp(\boldsymbol{\beta}_m^T \boldsymbol{x}_i)\exp\left[-\sum_{m=1}^{M}\int_0^{t_i}\gamma_m^b(\tau)\exp(\boldsymbol{\beta}_m^T \boldsymbol{x}_i)d\tau\right]$ is the discharge disposition specific likelihood function and $I_m$ is the index set of all residents who are discharged to disposition $m$. Further, to address the estimation issue of nonparametric component, we use partial likelihood maximization first to estimate $\boldsymbol{\beta}_m$, e.g., $\widehat{\boldsymbol{\beta}}_m = \arg\max_{\boldsymbol{\beta}_m} L_m(\boldsymbol{\Theta}_m \mid \boldsymbol{D})$, and then employ Efron morris estimator [25] to estimate $\gamma_m^b(\cdot)$.

## B. Service Need Heterogeneity Characterization

With the above quantified individual heterogeneity of NH LOS, we further investigate the individual heterogeneity of daily service need among NH residents during their NH stay. In real practice, the daily service need of NH residents can be quantified as per diem staff-time needed (in minutes) on each type of caregivers (e.g., CNA). Due to the diverse chronic conditions (e.g., vascular disease, osteoporosis, dementia and depression), multifunctional (e.g., physical and cognitive) limitations, and different types of therapies (e.g., audiology, occupational and/or physical therapy) as well as various treatments (e.g., radiation, dialysis and/or skin treatment) received, residents service need can be highly heterogeneous. For example, per diem CNA staff-time varies from 30 minutes to 200 minutes among different NH residents with varied health conditions and therapy service level. To characterize the heterogeneous service need, we develop a service need modeling framework, as shown in Figure. 2 (a). In this work, we particularly use CNA service need as an example to illustrate the model because CNAs compose the majority (over 60% in average) of nurse staff in Florida nursing facilities. Without losing generality, the proposed approach can be applied to any other types of caregivers in a NH as well (e.g., RN, LPN). In the proposed model, we incorporate NH domain knowledge, such as national Staff Time and Resource Intensity Verification (STRIVE) Project [9] and resource utilization group (RUGs) [26]. The raw RUG system is a patient classification system for reimbursement purpose which categorizes NH residents into multiple service need groups. Each service need group comprises NH residents with similar resource usage level. For residents in each of the raw service need groups, STRIVE project, one of the most recent national staff-time projects, further provides staff-time information based on nationwide measurement study. Specifically, STRIVE provides nationwide reference values of average daily staff-time needed based on the aggregation of raw staff-time measurements (collected using personal digital assistants) of approximately 97,000 NH residents from more than 200 representative high-quality NHs across different states. For each service need group, STRIVE contains (i) daily average staff-time spent directly with or on behalf of a resident; and (ii) daily average staff-time proportion spent indirectly for supporting the delivery of care for a resident. The former is defined as direct care staff-time, which involves activities such as feeding, helping dress, giving medications, charting for a resident, calling a physician about a resident, etc. The latter is defined as indirect care staff-time, which involves activities such as stocking medication cabinet, performing administration, participating in training sessions, taking time for breaks and meals, etc. Due to a lack of actual staff-time measurements at our studied nursing facility, we incorporate STRIVE project to quantify the representative daily staff-time needed (measured in minutes). We utilize the identifying variables in original RUGs classification system, as shown in Figure. 2 (b), and develop a modified service need classification system to characterize service need heterogeneity of NH residents. The identifying variables include (i) the resident-level ADL scores which measure the level of functional assistance or support required by the resident, and (ii) care intensity such as the rehabilitation service level and extensive medical care service level. The details are described as follows.

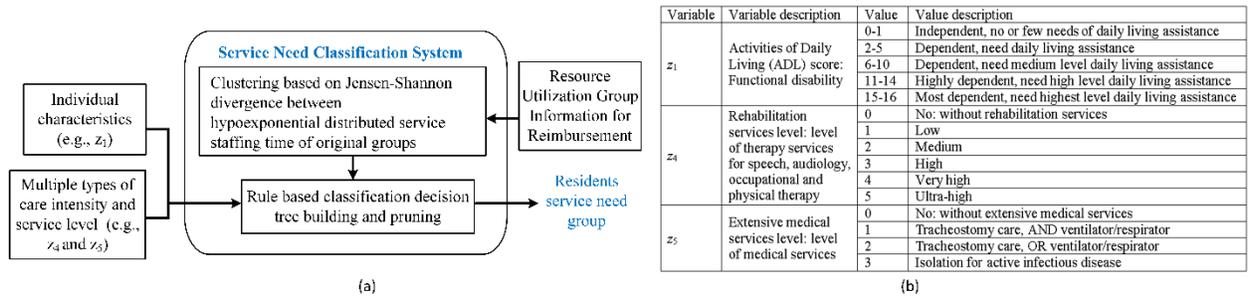

Fig. 2: (a) Service need heterogeneity characterization system, (b) identifying variables in the service need characterization system

*1) Staff-time Clustering:* As aforementioned, the raw RUG classification system is designed for reimbursement policy making. Consequently, NH residents in different raw groups may have similar patterns of daily service need of different types of caregivers. To reduce the complexity of raw classification system with no loss on model performance and to characterize service need heterogeneity of NH residents, we modify the raw RUG classification system. To ensure modeling simplicity and practical convenience [27], we consider exponential distribution with single parameter to model daily staff-time of direct care and indirect care, and further assume they are independent. The total staff-time then follows hypoexponential distribution with two different rate parameters. To capture similar patterns of total staff-time among NH residents, we perform clustering on total staff-time of original RUG groups. Specifically, we employ hierarchical clustering approach [28] and use Jensen-Shannon divergence (JSD) as similarity measure [29]. The number of target service need groups is obtained by minimum number of groups which can satisfy the condition that maximum within-cluster JSD is not exceeding tolerance $\epsilon_c$, expressed as $N^* = \arg\min N \in Z^+ : \left\| \left\{ D_{p_i, p_j} \right\}_{i,j \in c_k} \right\| \leq \epsilon_c, \forall k = 1, \dots, N$ where $D_{p_i, p_j} = -\int \frac{p_i + p_j}{2} \log \frac{p_i + p_j}{2} du + \frac{1}{2} \int p_j \log p_j du + \frac{1}{2} \int p_i \log p_i du$ with $p_i$ and $p_j$ as distributions of average daily staff-time of original RUG group $i$ and $j$, and $c_k$ is a index set of original groups which belong to cluster $k$. In our study, we use $\epsilon_c = 0.002$ as the criterion to determine final number of service need groups. Further, we obtain the group-specific parameters for total staff-time based on the mean of the rate parameters of all raw RUG groups within cluster. We evaluate the effectiveness of clustering results via Cramer-Von Mises (CVM) test [30]. The *p*-values of all groups based on CVM test are larger than 0.5. This strongly indicates that the hypoexponential distribution with fitted group-specific parameters achieves satisfactory goodness-of-fit performance. As shown in Figure. 3 (a), the total mean staff-time is increasing when the group index becomes larger and the distributions of total staff-time among different groups are significantly different.

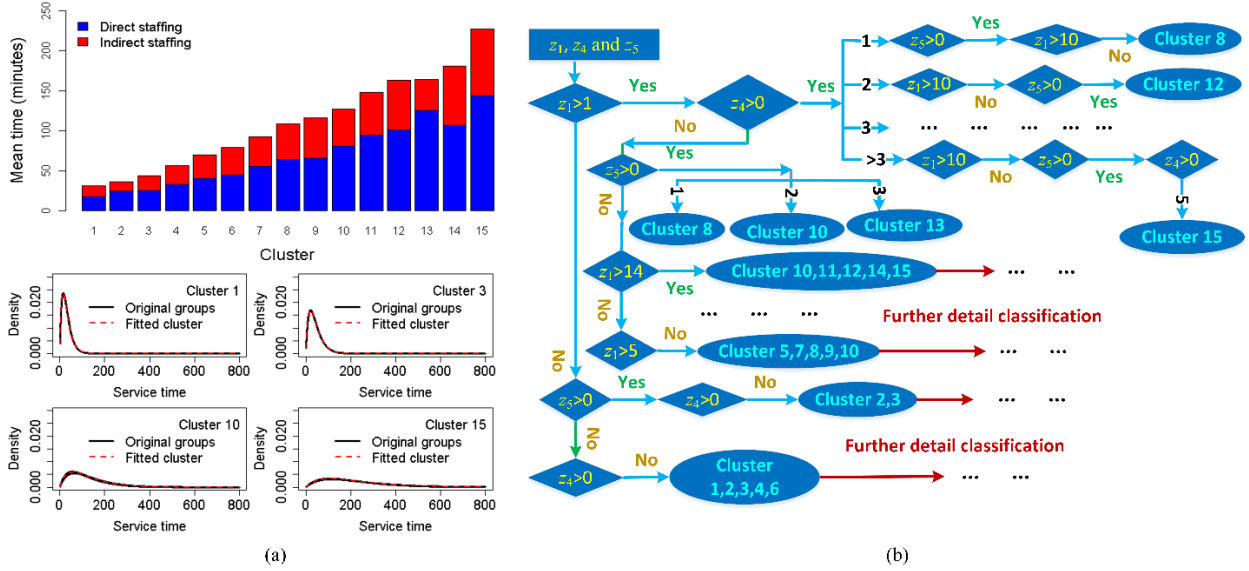

Fig. 3: (a) fitting performance of clusters, (b) classification decision tree

*2) Service Need Classification:* To further characterize the service need heterogeneity among different NH residents, we employ Apriori algorithm [31] to learn the association rules among identifying variables and service need clusters. Further, we construct and prune a classification tree [32]. The tree-based classification diagram is illustrated in Figure. 3 (b). When the resident ADL score is larger, the service need group index becomes larger, which implies that the residents need of CAN service becomes larger. It is because, in a typical NH, CNAs are mainly responsible for providing daily living assistance (e.g., dressing, feeding and bathing) for NH residents. Besides, CNAs also assist residents with their rehabilitation plans, such as assisting residents with physical or occupational therapy activities established by the therapists. As shown in Figure. 3 (b), when resident has higher rehabilitation service level, the CNA service need tends to become larger. In addition, when resident extensive medical care becomes more intensive, the resident need more CNA service, as depicted in Figure. 3 (b). It is because that CNAs also provide tracheostomy care and ventilator care (e.g., ventilation equipment setup) under verification and supervision from RN or LPN. Overall, the proposed classification system can characterize service need heterogeneity of NH residents with no loss on fitting performance and can ensure the modeling fidelity for demand simulation.

## C. Predictive Analytics Integrated Demand Simulator

With the above developed predictive models, we further integrate computer simulation to predict facility-level service demand of a heterogeneous population of NH residents over time for NH resource planning. In the predictive analytics integrated simulation model, we simulate daily arrivals for short-stayers and long-stayers. For each newly admitted short-stay resident, we employ the developed predictive LOS model to predict individual NH LOS and the discharge disposition. Due to the limited number of observations of discharge events for long-stay residents, we use distribution-based model to predict NH LOS for newly admitted long-stayers. For each NH resident, we identify individual service need group based on the developed classification system. For each resident within the identified service need group $g$, we extract group-specific daily average staff-time of direct care and indirect care, denoted as $y_g^1$ and $y_g^2$ respectively. The group-specific rate parameters of total staff-time then become $\lambda_g^s = \frac{1}{y_g^s}, s = 1,2$ with $s = 1$ for direct care and $s = 2$ for indirect care. For each resident $i$ who belongs to service need group $g$, denoted as $I_i = g$, we further simulate resident-level service demand $\xi_i$ on a daily basis measured by total staff-time in minutes, i.e., $\xi_i \mid I_i = g \sim HypoExp(\lambda_g^1, \lambda_g^2)$. Such resident-level daily service demand is repeatedly simulated for each NH resident until the end of his/her NH stay. Further, we aggregate the simulated resident-level service demand to obtain facility-level service demand of a heterogeneous population of N NH residents, denoted as $\xi$, i.e., $\xi = \sum_{i=1}^{N} \xi_i$. The generated facility-level service demand will then be utilized for NH resource planning.

*D. Two-phase Data-driven Optimization For NH Resource Planning*

To meet with the characterized service demand, we further develop a two-phase decision analytics module to determine optimal resource planning decisions at different time scales, such as annual bed capacity at strategic level, bi-monthly number of nursing staff to be recruited at tactical level, and daily working schedules as well as daily staffing level at operational level. The details are elaborated as follows.

*1) Capacity Planning*: In the first phase, the goal of capacity planning is to search for the optimal bed capacity such that the service quality measure of interest can be ensured. In this work, we consider average daily acceptance level as NH service quality measure. The optimality principle of capacity planning can be expressed as

$$\kappa_p = \arg\min\{\kappa \in \mathbf{Z}^+ : Pr(\bar{\alpha}_T(\kappa) \geq \tau_c) \geq \eta\} \quad (3)$$

where $\kappa_p$ is the minimum required bed capacity and $\bar{\alpha}_T(\kappa)$ is the mean of a vector of daily acceptance level over the planning horizon $T$ with the capacity fixed at $\kappa$, i.e., $\alpha_T(\kappa) = \{\alpha_t(\kappa)\}_{t=1}^T$. $\tau_c$ is service quality criterion and $\eta$ is the quantile threshold. To account for the uncertainty of residents flow and to achieve desired facility-level performance, we employ simulation-based heuristic search to determine the minimum

---

**Algorithm 1** Optimal Capacity Planning

**Initialization:** $\kappa_p \leftarrow \kappa_0$, $\zeta \leftarrow 0.5\kappa_0$, $m \leftarrow 1$, $\upsilon \leftarrow 1$, $N \leftarrow 500$, $r \leftarrow 500$, $T \leftarrow 365$, $\tau_c \leftarrow 85\%$, $\eta \leftarrow 95\%$

**while** $m \leq N \& \upsilon = 1$ **do**
    **for all** $j \leftarrow 1$ to $r$ **do**
        **for all** $t \leftarrow 1$ to $T$ **do**
            obtain daily acceptance level $\alpha_t^j(\kappa_p)$ with current capacity $\kappa_p$ at day $t$ of run $j$ in iteration $m$, $\forall t = 1...T$
        **end for**
        obtain average daily acceptance rate in run $j$ as:
        $\bar{g}_j = \frac{1}{T} \sum_{t=1}^{T} \alpha_t^j(\kappa_p)$
    **end for**
    calculate $p = Pr(\bar{g} \geq \tau_c)$
    **if** $p < \eta$ **then**
        update $\zeta \leftarrow 0.5\kappa_p$ and $\kappa_p = \lceil \kappa_p + \zeta * (\eta - p) \rceil$
    **else**
        $\upsilon = 0$
    **end if**
    $m \leftarrow m + 1$
**end while**

---

required number of beds at the desired service quality level. The optimal capacity algorithm is summarized in Algorithm. 1. In each iteration, we perform 500 replication runs to simulate NH arrivals, admission and discharge events, and then obtain 500 samples of daily average acceptance rate based on simulation outputs. We then check if current capacity decision achieves the satisfactory accessibility performance. If the stopping condition is not satisfied, we will update current bed capacity and evaluate the performance with a new capacity decision. The iterative searching process continues until we find the optimal capacity that can achieve the desired performance criterion. The developed solution algorithm can maximize the utilization performance (e.g., bed occupancy) while achieve a desired level of care accessibility (e.g., average daily acceptance rate at 85% or above).

*2) Workforce Planning*: With the simulated facility-level service demand and the determined optimal annual capacity, we further formulate an integrated staffing and scheduling (ISS) optimization model to

meet with the heterogeneous service demand of NH residents at minimized overall labor cost. To reduce

---

**Algorithm 2** Optimal Workforce Planning

**function** PATTNGEN($x$, $\Gamma$, $t_c$, $T_s$, $\eta$, $v_{pf}$, $\tilde{t}_1$, $\tilde{t}_2$)
    **if** $|\eta| = t_c$ **then**
        $\Gamma \leftarrow \Gamma \cup \eta$
    **else**
        **for** $j \leftarrow x$ to $T_s$ **do**
            check weekend preference, $c \leftarrow$ total number of working days allocated at weekend
            **if** $c \leq \tilde{t}_1$ **then**
                check part time constraints, $c \leftarrow$ bi-weekly total number of working days
                **if** $v_{pf}=2$ AND $c \leq \tilde{t}_2$ OR $v_{pf} = 1$ **then**
                      $\eta \leftarrow \eta \cup j$
                      PATTNGEN($j$+1, $\Gamma$, $t_c$, $T_s$, $\eta$, $v_{pf}$, $\tilde{t}_1$, $\tilde{t}_2$)
                **end if**
            **end if**
        **end for**
    **end if**
**end function**

**procedure** STAFFDECISIONMAKER
    PATTNGEN(1, $A$, $\beta_{ft}$, $T$, $\Xi$, 1, $\tilde{t}_{wk}$, $\tilde{t}_{pt}$)
    **for each** $\beta \in \beta_{pt}$ **do**
        $\Xi \leftarrow \{\}$
        PATTNGEN(1, $A$, $\beta$, $T$, $\Xi$, 2, $\tilde{t}_{wk}$, $\tilde{t}_{pt}$)
    **end for**
    simulate individual daily service demand of NH residents
    aggregate resident-level demand to obtain facility-level demand as $\xi = [\xi_1, ..., \xi_T]^T$
    solve the following model by sample average approximation:
    $\min_{x \subseteq \mathbb{Z}_+^P} \; c^T x - c_u \mathbb{E}_{\xi}\left[[(s - \xi)^-]^T \mathbb{1}_T\right] + c_v \mathbb{E}_{\xi}\left[[(s - \xi)^+]^T \mathbb{1}_T\right]$
    where $(d)^- = [\min(d_1, 0), ..., \min(d_T, 0)]^T$ and $(d)^+ = [\max(d_1, 0), ..., \max(d_T, 0)]^T$
    $s = [s_1, s_2, ..., s_T]^T$ and $s_t = K\left(\sum_{i=1}^P A_{it} x_i\right)$
**end procedure**

---

the combinatorial complexity of working schedules in real practice, we apply dynamic programming to generate the scheduling patterns based on NH industrial knowledge, as described by *PATTNGEN* in Algorithm. 2. $A_{P \times T}$ is a matrix of all generated scheduling patterns. $A_{it} = 1$ if scheduling pattern $i$ contains working day $t$, and $A_{it} = 0$ otherwise. The objective of workforce planning is to achieve optimal staffing and scheduling decision at reduced total labor cost. The total labor cost includes planned staffing cost for staff recruitment, understaffing penalty cost for calling temporary nursing staff (e.g., nurse as needed or agency aides) to satisfy the unmet demand, and overstaffing penalty due to unnecessary planned additional staffing. We denote the understaffing penalty cost as $c_u$ and denote overstaffing penalty as $c_v$. We use $x = [x_1, ..., x_P]^T$ to represent a vector of decision variables where $x_i$ is number of staff need to be recruited for scheduling pattern $i$, and use $c = [c_1, ..., c_P]^T$ to represent a vector of staffing cost where $c_i$ is the staffing cost of scheduling pattern $i$, $\forall i = 1, ..., P$. Given the planning horizon $T$, we can then define a loss function as $L(\xi, x) = c^T x - c_u \sum_{t=1}^T (s_t - \xi_t)^- + c_v \sum_{t=1}^T (s_t - \xi_t)^+$ where the operators $(\cdot)^-$ and $(\cdot)^+$ are denoted as $(d)^- = \min(0, d)$ and $(d)^+ = \max(0, d)$. $\xi_t$ refers to the random facility-level service demand at day $t$, $\forall t = 1, ..., T$. $s_t$ refers to the staffing supply in minutes at day $t$ and can be further expressed as $s_t = K(\sum_{i=1}^P A_{it} x_i)$, where $P$ is total number of generated scheduling patterns and $K$ is the daily supply (in minutes) per staff. Then the goal of NH workforce planning is to determine number of nursing staff need to be recruited for each scheduling pattern that can minimize the expected loss function and can meet with facility-level service demand as well. To account for the stochastic uncertainty of service demand, we employ sample average approximation method [33] to solve the ISS optimization model. The NH ISS optimization problem and solution procedure are summarized in Algorithm. 2.

*E. Implementation*

With the above proposed modeling approaches, we further develop a practical user-friendly decision support platform to facilitate managerial decision making for NH administrators. The predictive analytics integrated simulation module for service demand characterization and the decision analytics module for resource planning are developed using C++ in Visual Studio environment. The user interface of the decision support tool is illustrated in Figure. 4. This analytics-based data-driven decision support system for NH resource planning can deliver rich decisions at different time scales, including strategic decision, such as annual bed capacity decision, tactical decision, such as bi-monthly staff recruitment decision (e.g., number of nursing staff need to be recruited), and operational decision, such as daily staffing levels as well as daily working schedules.

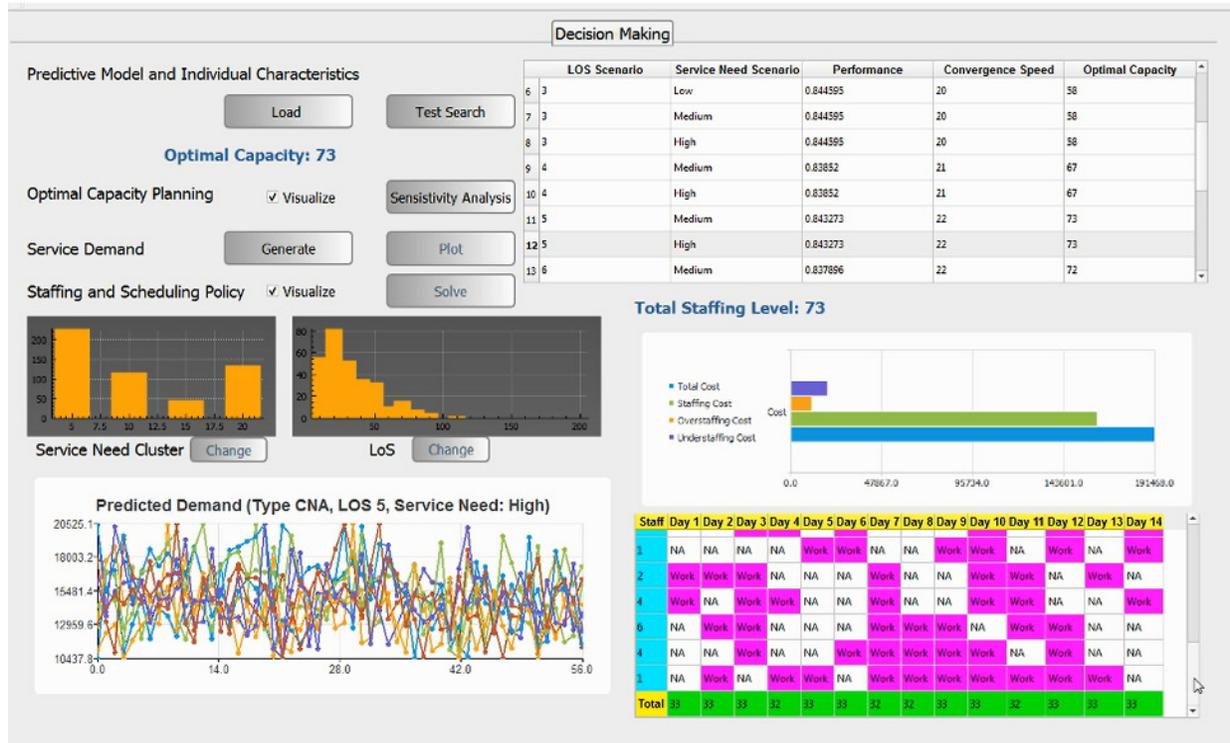

Fig. 4: Graphical user interface of the decision support platform

## IV. CASE STUDY

*A. Data Description*

To evaluate the prediction performance and decision performance of the proposed framework, we use the de-identified electronic health records of NH residents [34] from our industrial collaborator (Greystone Healthcare Management Corp.). The real data contains details of admission and discharge records, and rich resident-level health assessment information, including but not limited to socio-demographics, clinical diagnoses, chronic conditions, functional performances limitations (e.g., physical limitation and cognitive impairments, etc.), and care intensity as well as service level. In this work, totally 677 residents are considered. Among the residents, the majority are elderly adults over 65 years old. Most of the aging residents are frail and many of them have multiple chronic conditions. Over 90% of the NH residents have at least one type of chronic diseases (e.g., cancer, and hypertension, etc.). About 38% of residents have diabetes mellitus and about 35% of residents have anemia condition. In addition, about 5% of residents are diagnosed with acute conditions such as obstructive uropathy. Further, the ADL score of NH residents are also available from the data set. It measures the level of functional assistance or support required by the residents. A higher ADL score means that the resident is more physically dependent and needs more daily living assistance. In the real data, over 90% of residents need daily living assistance (e.g., ADL is larger than 1). Among the residents who need daily living assistance, 20% of them have more physical disabilities and are highly dependent, and thus need most daily living assistance (e.g., ADL from 11 to 16). Besides, the resident-level information also includes the varied service need level and care intensity among different types of services. Over 95% residents need rehabilitation therapy service during their stay. On the other hand, less than 5% residents need extensive medical care service. Moreover, based on the admission and discharge records, the resident-level LOS information is extracted as well.

*B. Prediction Performance Evaluation and Comparison*

Based on the acquired data, we first model simulation inputs, such as daily arrivals and LOSs of NH residents, and evaluate their prediction performances. Based on the LOS observations and definitions of short stay as well as long stay from Centers for Medicare & Medicaid Services, we classify NH residents into short-stay residents (e.g., LOS ≤ 100 days) and long-stay residents (e.g., LOS > 100 days). Short-stay residents mainly receive rehabilitation therapy service and post acute care while long-stayers mainly receive long-term custodial care in a NH. Negative binomial distribution is used to model the arrivals of short-stay residents, i.e., NB($r, p$). We evaluate the goodness-of-fit performance by Chi-square test. The $p$-value of the estimated NB($\hat{r}, \hat{p}$) is 0.3, which indicates satisfactory goodness-of-fit of real arrival data. On the other side, we use Poisson distribution to model the arrivals of long-stay resident, i.e., Pois($\lambda$). The $p$-value of Chi-square test is 0.67, which also indicates a satisfactory goodness-of-fit performance.

Further, we apply the developed LOS model to analyze the real LOS observations. About 61% of residents are discharged to community and 24% of residents are re/hospitalized. Thus, we consider these two major discharge dispositions, e.g., community and hospital. Other discharge dispositions, such as transferring to another NH or death, are negligible and thus discarded. In the real NH data, more than 90% of residents are short-stay residents (e.g., LOS ≤ 100 days) while the others stay longer than 100 days. We employ the developed predictive LOS model with incorporating individual characteristics and multiple competing discharge dispositions to analyze NH LOS of short-stay residents. We employ variable selection techniques and identify 16 covariates to model the heterogeneous LOS, including both time-to-community and time-to-hospital. As shown in Figure. 5, the ADL score exhibits a significant negative effect on the time-to-community. With a higher ADL score, the resident needs more daily living assistance and thus stay in NH longer. It is because the resident either needs more time to restore his/her functional performance, or changes from short-term rehabilitation stay into long-term care stay due to unsuccessful recovery. The proposed LOS model also identifies other significant factors, including varied disease conditions, such as anemia, diabetes, and obstructive uropathy, etc. These factors have significant positive effects on rehospitalization risk, as shown in Figure. 5. The quantified effects of these contributing factors on re/hospitalization risk can better inform NH administrators to focus on the most at-risk NH residents with more targeted resource allocated.

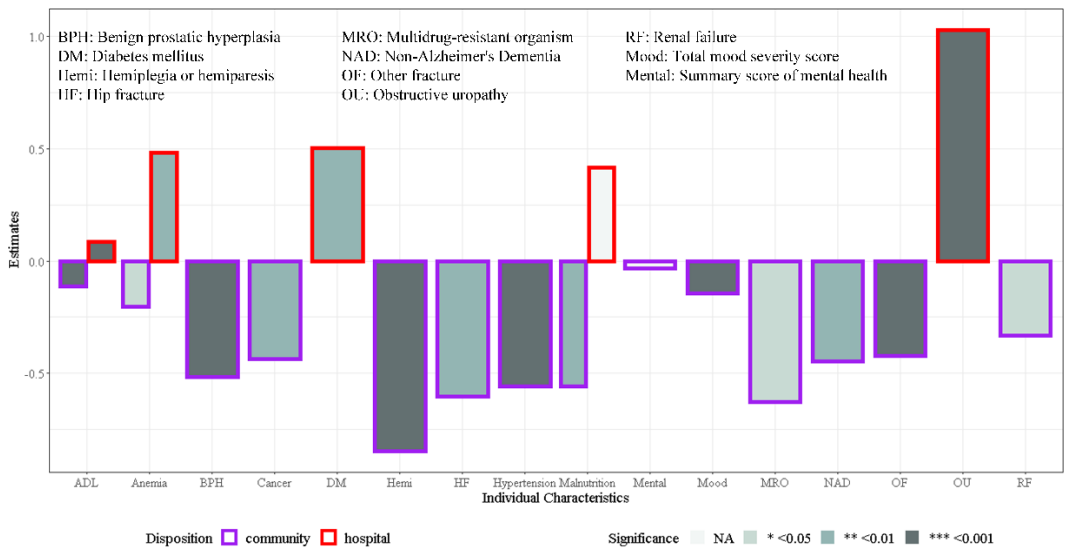

Fig. 5: Estimated effects of disposition-specific influencing factors

Based on the developed model, we further evaluate its prediction performance and compare with alternative modeling approaches, namely LOS model without considering multiple discharge dispositions.

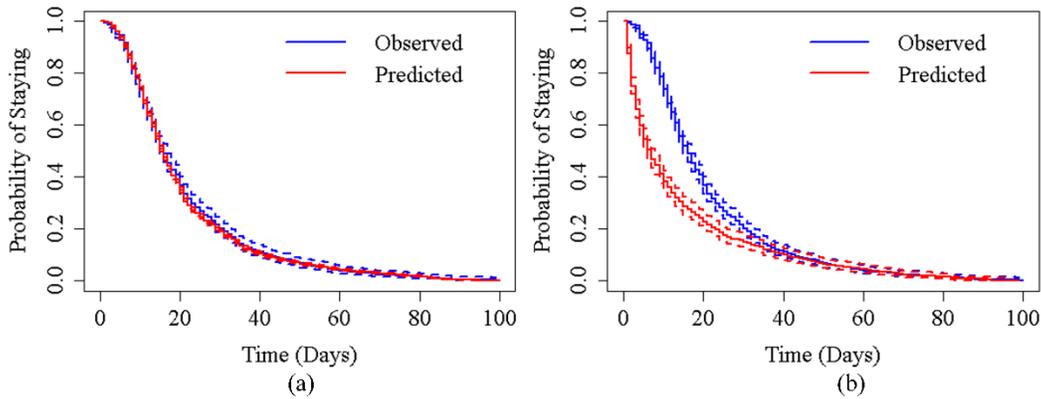

Fig. 6: Comparison between the observed (blue) and the predicted (red) survival curves based on different modeling approaches, (a) proposed model, (b) alternative model

Specifically, we compare the Kaplan-Meier survival curves [36] between predicted samples of two LOS models and observed samples. As shown in Figure. 6(a), the K-M curve of predicted LOS samples based on the proposed model is close to the K-M curve of observed LOS samples. This demonstrates satisfactory predictive distribution accuracy of the developed LOS model. On the other side, the model without considering multiple discharge dispositions results in LOS underestimation, as shown in Figure. 6(b). Overall, the developed LOS model can successfully capture LOS heterogeneity with multiple competing discharge dispositions and improve LOS prediction performance. For long-stay residents, most of them stay much longer than the selected one-year period. Thus, few completed discharge events are observed. We simply use log-normal distribution to capture such right-censored observations. The *p*-value of Chi-square test is 0.26 and thus the estimated model of LOSs for long-stay residents is also validated.

## C. Predictive Analytics Integrated Simulation Output Validation

With the above predictive analytics, we further integrate them with high-fidelity simulation to predict facility-level service demand of NH residents. To validate the simulation model, we compare the simulated outputs of daily resident volume with the actual observed resident volume. As shown in Figure. 7 (a), the simulated samples of daily resident volume exhibit a similar distribution to the real data. Moreover, we use two-sample Kolmogorov-Smirnov (KS) test to compare their statistical differences and a *p*-value of 0.52 implies that there is no statistically significant difference between the two. The results demonstrate the superior demand prediction performance of the proposed approach. On the other hand, the simulation outputs of alternative model without considering competing discharge dispositions fail to achieve satisfactory prediction results, as shown in Figure 7 (b), with a *p*-value of <2.2e-16 in the KS test. The conventional model which neglects multiple discharge dispositions will result in LOS underestimation and a lower-than-actual resident volume. Such prediction inaccuracy will propagate to resource planning, and further induce the inappropriate capacity planning and ISS decisions with unsatisfactory performance (e.g., inadequate capacity, and large understaffing cost). To further provide validation of the simulation outputs at a finer scale, we compared the simulated daily number of residents over time (under multiple replication runs) based on the proposed approach with the real data. As shown in Figure. 7 (c), 95% confidence bands of the simulated resident volume can fully cover the observed daily number of residents over time. On the other hand, the simulation outputs based on alternative LOS model without considering multiple competing discharge dispositions have lower-than-actual daily number of residents over time than the actual data, as shown in Figure. 7 (d).

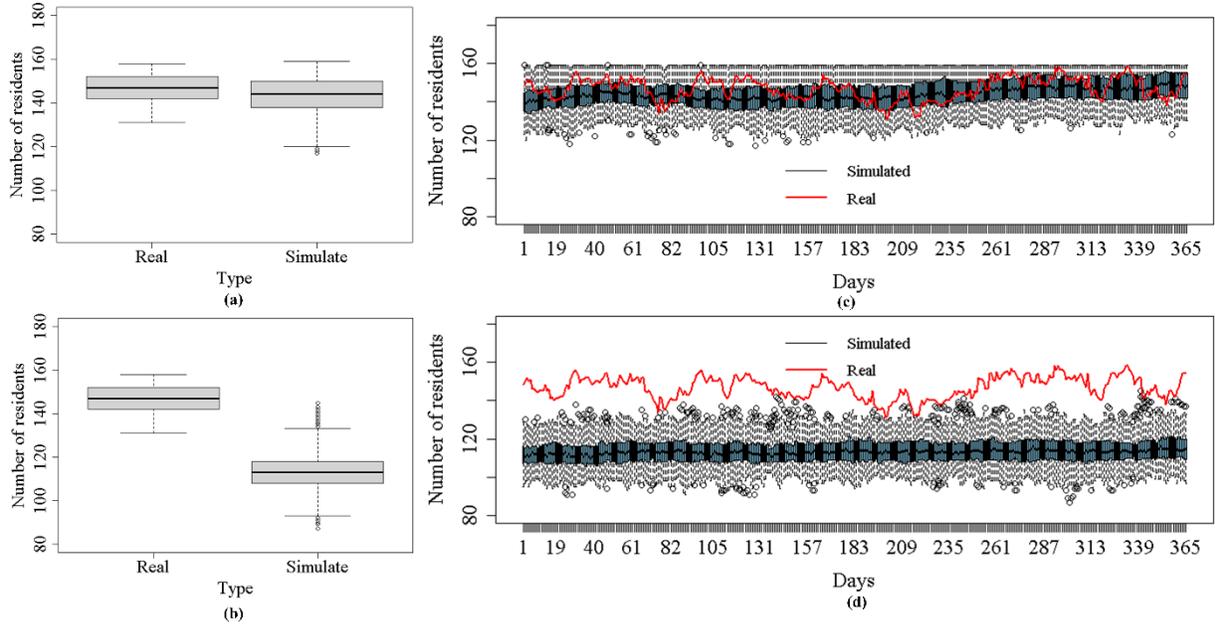

Fig. 7: Comparison between observed data and simulated samples of daily number of residents as well as simulated outputs at finer scale, (a)(c) based on proposed model, (b)(d) based on alternative model

Further, we validate the developed simulation model by comparing the simulated and observed system performance (e.g., daily staffing level, average planned staffing cost per day). Specifically, we fix same capacity as the number of beds in real practice to simulate the facility-level service demand and use the SR ratio based staffing strategy without scheduling patterns, which is implemented in real nursing facility. The comparison between simulation based result and real world practice is summarized in Table. I. The average planned staffing cost per day and the daily staffing level from the proposed model are covered by the lower and upper bounds of observed performance. This further justifies the validity of the proposed predictive analytics integrated simulation model for demand simulation.

TABLE I: System performance comparison of real practice and proposed model

| Cost Analysis | Real Practice | | | Simulation | |
|---|---|---|---|---|---|
| | Min | Mean | Max | Lower bound | Upper bound |
| Average planned staffing cost per day | 2.9K | 4.2K | 5.2K | 3168 | 4224 |
| Daily staffing level | 32 | 46 | 57 | 36 | 48 |

*D. Decision Performance Evaluation and Comparison*

With the characterized service demand from the proposed predictive analytics integrated high-fidelity simulation, we further implement the proposed two-phase optimization module for NH resource planning. For capacity planning, we employ heuristic search method to find the optimal bed capacity with the desired performance level, as described in Section. III-D1. To evaluate the capacity decision performance, we compare the proposed capacity strategy with other alternative strategies, as shown in Table. II. As compared to the other capacity strategies (e.g., C1-C4), the proposed capacity strategy can achieve the desired accessibility level (e.g., average daily acceptance rate 85% or above) at largest bed utilization performance with minimum required number of beds, as shown in Figure. 8. On the other side, the other strategies either fail to achieve the desired quality level of care accessibility (e.g., C2 and C4), or fail to maintain a satisfactory bed utilization performance (e.g., C1 and C3). Moreover, we evaluate the performance of capacity decision from the alternative LOS model which neglects multiple competing discharge dispositions. The required number of beds derived from the LOS model without considering multiple discharge dispositions is 100, which is much smaller as compared to the capacity from propose strategy due to the underestimated LOS and lower resident volume. The average daily acceptance rate based on such capacity decision becomes less than 30% and it fails to achieve the desired care accessibility level. Overall, the capacity obtained from the proposed strategy can achieve both of the desired accessibility level and bed utilization performance with the minimum required number of beds.

TABLE II: Descriptions of different capacity strategies

| Strategies | Description |
|---|---|
| C1 | The maximum capacity of NHs in the state of Florida |
| C2 | The average capacity of NHs in the state of Florida |
| C3 | The maximum capacity of NHs from our industry collaborator |
| C4 | The average capacity of NHs from our industry collaborator |

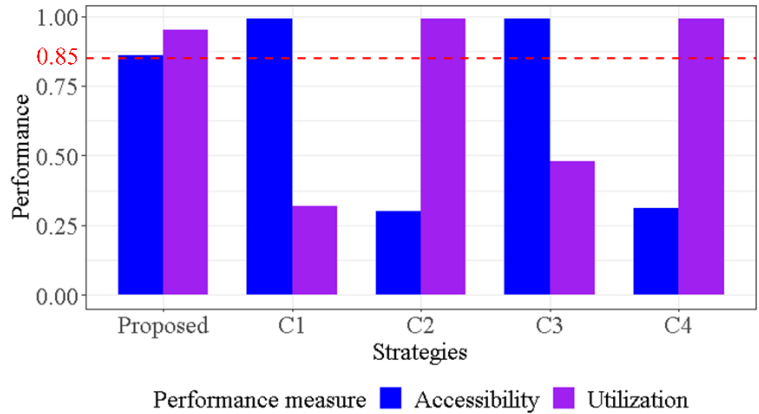

Fig. 8: Performance comparison among different capacity strategies

With the optimal annual capacity obtained, we further evaluate the performance of proposed ISS decision making model. Without losing generality, we consider workforce planning decision of CNAs since CNAs provide the most direct and essential care to the NH residents. We implement Algorithm. 2 and solve the ISS optimization model by sample average approximation method. We empirically investigate the variability of total labor cost calculated from derived staffing decision over different sample size. As shown in Figure. 9, the solution procedure can achieve stability and convergence when sample size is larger than 70. To account for both computational complexity and convergence, we choose 100 as the selected sample size. With the obtained staffing decision, we then compare with other alternative staffing strategies, as described in Table. III. As compared to the other staffing strategies (e.g., M1-M6), the proposed ISS strategy achieves the smallest total labor cost, as shown in Figure. 10, due to the following three reasons. First, the proposed strategy accounts for the heterogeneous service need of NH residents while the SR ratio-based strategies (e.g., M1-M4) fail to consider the service need heterogeneity. The strategies with minimum SR ratio elicited from industry practice (e.g., M1 and M3) result in larger overstaffing penalty. On the other side, the strategies with minimum SR ratio enforced by state regulation (e.g., M2 and M4) lead to significantly larger understaffing cost. Besides, as compared to the SR ratio-based staffing strategies without considering scheduling patterns (e.g., M1 and M2), the SR ratio-based ISS strategies (e.g., M3 and M4) which incorporate scheduling patterns can achieve smaller total labor cost. All of these above strategies result in larger unmet service demand, and further induce larger total cost than the proposed need-based ISS strategy. Second, the proposed strategy accounts for the heterogeneous LOS of NH residents via the developed LOS predictive model. We compare the proposed strategy with other need-based strategy (e.g., M5) without incorporating the improved LOS predictive analytics and further investigate the impact of LOS heterogeneity on ISS decisions. Due to the underestimated LOS and the smaller bed capacity, the strategy M5 underestimates the service demand, and consequently results in significantly larger understaffing cost than the proposed strategy, as shown in Figure. 10. Third, the proposed strategy accounts for the stochastic uncertainty of service demand via stochastic optimization. We compare the proposed strategy with other need-based strategy (e.g., M6) based on deterministic optimization to investigate the impact of stochastic demand uncertainty on ISS decisions. The strategy M6 fails to consider service demand uncertainty and leads to larger total labor cost as compared to the proposed strategy. The value of stochastic solution is 242. Overall, the proposed need-based ISS strategy can successfully meet with the heterogeneous service demand and achieve the smallest total labor cost.

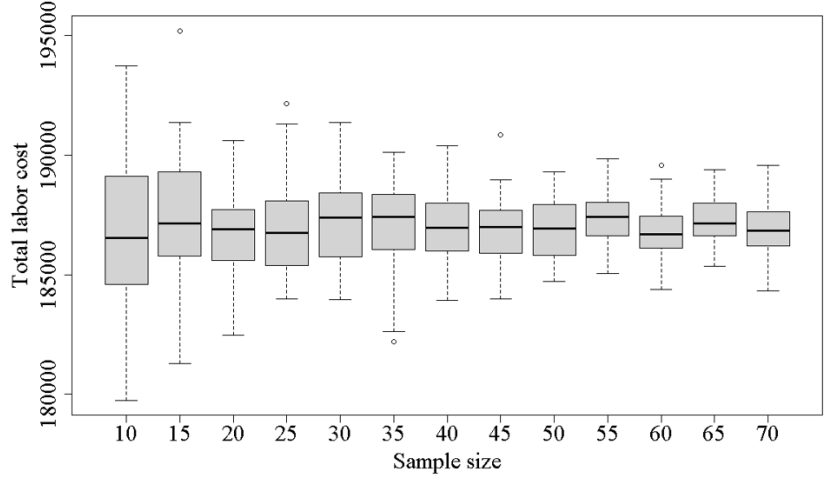

Fig. 9: Variability of total labor cost over different sample size

TABLE III: Descriptions of different staffing strategies

| Strategies | Description |
|---|---|
| M1 | Facility-implemented SR ratio-based staffing without scheduling patterns |
| M2 | State-regulated SR ratio-based staffing without scheduling patterns |
| M3 | Facility-implemented SR ratio-based ISS |
| M4 | State-regulated SR ratio-based ISS |
| M5 | Need-based ISS without incorporating predictive analytics |
| M6 | Need-based ISS without incorporating stochastic uncertainty of demand |

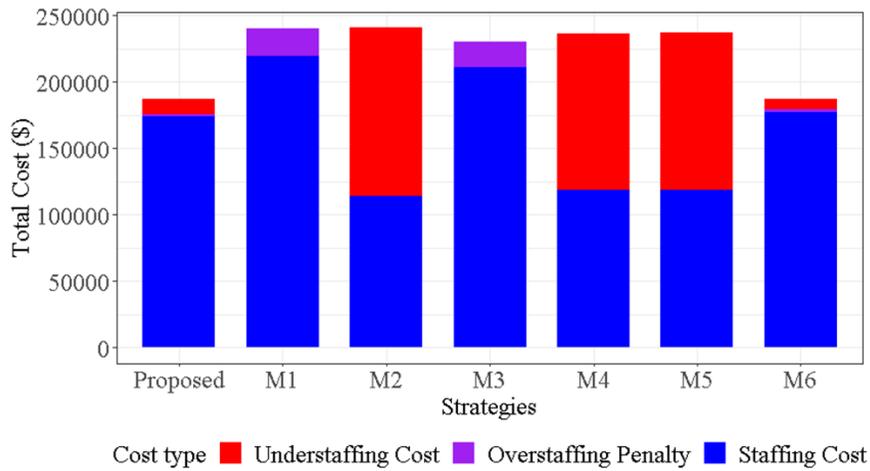

Fig. 10: Performance comparison among different staffing strategies

*E. What-if Scenarios with Different Resident Census Compositions*

Given the above resident composition, the proposed work demonstrates superior performances on both capacity and workforce planning decisions. In real practice, the current strategies may need to be adjusted when the resident composition varies considerably, such as in the same NH but at different seasonal/yearly period or across different NH facilities. Thus, we generate different what-if scenarios of resident census composition to investigate how the resource planning decisions may vary. Particularly, we investigate the impacts of LOS heterogeneity and service need heterogeneity on resource planning decisions. We consider the actual resident composition scenario in Section. IV-A as baseline scenario (S1). We then generate five more alternative scenarios, including

- worse health conditions scenario (S2): as compared to S1, half of residents have diabetes mellitus and the percentage of residents having anemia increases to 50%. Over 50% of residents have obstructive uropathy.
- less physically dependent scenario (S3): 90% of residents are less physically dependent (e.g., ADL from 0 to 1) and can either live independently or require less daily living assistance. The mean of ADLs decreases 80% as compared to S1.
- more physically dependent scenario (S4): 90% of residents are more physically dependent (e.g., ADL from 11 to 16) and require more daily living assistance. The mean of ADLs increases 80% as compared to S1.
- less rehabilitation therapy need scenario (S5): residents have the same ADLs and health conditions as those in baseline scenario S1. However, the percentage of residents who need rehabilitation decreases 50% as compared to S1.
- more extensive medical service need scenario (S6): as compared to S1, residents have the same ADLs, health conditions and therapy needs. However, the percentage of residents who require extensive medical care increases to 90%.

Based on the constructed scenarios, we investigate the effects of different census compositions on service demand and further investigate the effects of LOS heterogeneity and service need heterogeneity on capacity and workforce planning decisions. The performance comparison results among different scenarios are summarized in Table. IV. The resident-level service utilization quantified by LOS and service demand quantified by daily staff-time needed in minutes may vary in response to the varied resident census composition. Consequently, the capacity and workforce planning decisions will be affected by the varied service utilization and the varied service demand. For example, the residents in S2 have shorter NH stay than S1 due to the increased likelihood of being re/hospitalized. Such increased likelihood is attributed to the increased percentage of residents having acute conditions (e.g. anemia, diabetes mellitus and obstructive uropathy), which impose significantly positive effects on the re/hospitalization risk (as illustrated in Figure. 5). Consequently, the residents are more likely to be re/hospitalized more quickly for the acute condition treatment and thus have shorter NH stay. Due to shorter NH LOS, the resident volume decreases, and finally the optimal capacity in S2 decreases, as shown in Table. IV. Nevertheless, there is no change of service need in S2 because the identifying variables for service need group (e.g., individual functional performance, therapy intensity, etc.) are the same as compared to S1. With the joint effects of less residents staying in NH and unaltered daily service need, the facility-level service demand decreases as well, and further the planned staffing cost and total labor cost also decrease, as shown in Table. IV. In addition to the varied health conditions, the varied functional performances among different scenarios also affect residents NH LOS and daily service need, and further impact capacity and workforce planning decisions. For example, the residents in S3 have decreased ADL score and thus need less daily living assistance. With the joint effects of increased likelihood of being discharged to community and decreased likelihood of re/hospitalization (as illustrated in Figure. 5), the residents in S3 tend to stay shorter in NH and the resident volume decreases accordingly. Consequently, the optimal capacity in S3 decreases as compared to S1. The daily service need also decreases due to less care services are needed. Due to both shorter stay and less service need, the facility-level service demand decreases and further the total labor cost as well as planned staffing cost in S3 decrease as well, as shown in Table. IV. On the other side, the residents in S4 become more physically dependent with larger ADL scores and thus need more daily living assistance during NH stay. The residents in S4 tend to stay longer because of the joint effects of decreased likelihood of being discharged to community and increased re/hospitalization risk. As a result, the optimal capacity in S4 increases significantly with increased resident volume. The daily service need also increases in response to the increased ADL scores. Due to the increased service utilization (measured in LOS) and the increased daily service need, the facility-level service demand increases in S4. Thus, more CNAs need to be recruited in S4 to meet with such increased facility-level demand as compared to S1. Consequently, the total labor cost and planned staffing cost in S4 become significantly larger, as shown in Table. IV. Moreover, the care intensity of different types of service also varies among different scenarios and induces varied service need, which further affects workforce planning decision. For example, the rehabilitation need of NH residents in S5 decreases 50% as compared to S1, and thus the service need becomes less. Nevertheless, the service utilization of NH residents (measured in LOS) in S5 are the same as S1, which is different from the varied service utilization among scenarios S2-S4. Thus, there is no difference in capacity decision of S5 as compared to S1 due to the unchanged resident volume. However, the planned staffing cost in S5 becomes smaller and the total labor cost decreases as compared to S1, as shown in Table. IV, due to the decreased service need. Fewer CNAs are needed to meet with the decreased service demand in S5. On the other side, the need for extensive medical care increases in S6, and consequently the characterized service need increases as compared to S1. Despite increased residents service need, there is no change of the individual characteristics related to service utilization in S6 as compared to S1. Thus, the optimal capacity in S6 is the same as the capacity decision in S1 due to the unaltered resident volume. Due to the increased daily service need and unchanged service utilization, the facility-level service demand increases. Consequently, more CNAs are needed to meet with the increased demand and thus the total labor cost as well as the planned staffing cost in S6 become larger as compared to S1, as shown in Table. IV. Overall, the proposed framework is flexible to analyze different what-if scenarios of resident composition and can suggest adaptive optimal resource planning decisions to NH administrators.

TABLE IV: Resource planning decision comparison among different scenarios

| Scenario | Optimal Capacity | Total Labor Cost ($) | Planned Staffing Cost ($) |
|---|---|---|---|
| S1 | 156 | 187007 | 173888 |
| S2 | 148 | 178607 | 167552 |
| S3 | 146 | 131249 | 123904 |
| S4 | 167 | 240451 | 224928 |
| S5 | 156 | 180247 | 167904 |
| S6 | 156 | 225031 | 208736 |

## V. CONCLUSION

In this chapter, we present a methodological innovation to resolve the complexities of characterizing heterogeneous service demand of NH residents and to improve the system performance of resource planning decisions to meet with such complex service demand. Specifically, we introduce an analytics-based modeling framework and decision support system to facilitate NH resource planning under heterogeneous service demand of NH residents. The novel integration of advanced predictive analytics, computer simulation, and applied stochastic optimization as well as domain knowledge in real practice addresses a series of data and decision complexity in NH resource preparedness and care delivery. The proposed predictive analytics integrated simulation is able to characterize the heterogeneous service demand among NH residents with higher granularity and allows accurate prediction of service demand. The developed decision analytics tools further provide optimal resource planning decisions with rich managerial insights. The derived operational policy for capacity and workforce planning is implementable and actionable. The proposed framework is beneficial to multiple stakeholders in NH industry, such as facilitating managerial decisions making of NH administrators, and improving health outcomes of NH residents. In the future work, we will investigate caregiver assignment decision under the developed framework and decision support system.

## REFERENCES

[1] H. Janiszewski Goodin, "The nursing shortage in the united states of america: an integrative review of the literature," Journal of advanced nursing, vol. 43, no. 4, pp. 335–343, 2003.
[2] K. Hyer, K. S. Thomas, L. G. Branch, J. S. Harman, C. E. Johnson, and R. Weech-Maldonado, "The influence of nurse staffing levels on quality of care in nursing homes," The Gerontologist, vol. 51, no. 5, pp. 610–616, 2011.
[3] K. Spilsbury, C. Hewitt, L. Stirk, and C. Bowman, "The relationship between nurse staffing and quality of care in nursing homes: a systematic review," International journal of nursing studies, vol. 48, no. 6, pp. 732–750, 2011.
[4] J. Park and S. C. Stearns, "Effects of state minimum staffing standards on nursing home staffing and quality of care," Health services research, vol. 44, no. 1, pp. 56–78, 2009.
[5] R. T. Konetzka, S. C. Stearns, and J. Park, "The staffing-outcomes relationship in nursing homes," Health services research, vol. 43, no. 3, pp. 1025–1042, 2008.
[6] National Center For Health Statistics, "Long-Term Care Providers and Services Users in the United States: Data From the National Study of Long-Term Care Providers," https://www.cdc.gov/nchs/npals, 2019, online; accessed 26 Jan 2021.
[7] K. G. Manton, K. Liu, and E. S. Cornelius, "An analysis of the heterogeneity of us nursing home patients," Journal of Gerontology, vol. 40, no. 1, pp. 34–46, 1985.
[8] K. Liu, T. Coughlin, and T. McBride, "Predicting nursing-home admission and length of stay: A duration analysis," Medical Care, pp. 125–141, 1991.
[9] Centers For Medicare & Medicaid Services, "Staff Time and Resource Intensity Verification (STRIVE)," https://www.cms.gov/Medicare/Medicare-Fee-for-Service-Payment/SNFPPS/TimeStudy, 2013, online; accessed 26 January 2021.
[10] C. Mueller, G. Arling, R. Kane, J. Bershadsky, D. Holland, and A. Joy, "Nursing home staffing standards: Their relationship to nurse staffing levels," The Gerontologist, vol. 46, no. 1, pp. 74–80, 2006.
[11] X. Zhang and D. C. Grabowski, "Nursing home staffing and quality under the nursing home reform act," The

Gerontologist, vol. 44, no. 1, pp. 13–23, 2004.

[12] J. R. Bowblis, "Staffing ratios and quality: An analysis of minimum direct care staffing requirements for nursing homes," Health services research, vol. 46, no. 5, pp. 1495–1516, 2011.

[13] J. R. Bowblis and A. Ghattas, "The impact of minimum quality standard regulations on nursing home staffing, quality, and exit decisions," Review of Industrial Organization, vol. 50, no. 1, pp. 43–68, 2017.

[14] J. Ridge, S. Jones, M. Nielsen, and A. Shahani, "Capacity planning for intensive care units," European journal of operational research, vol. 105, no. 2, pp. 346–355, 1998.

[15] A. Kunkel and L. A. McLay, "Determining minimum staffing levels during snowstorms using an integrated simulation, regression, and reliability model," Health care management science, vol. 16, no. 1, pp. 14–26, 2013.

[16] G. Ma and E. Demeulemeester, "A multilevel integrative approach to hospital case mix and capacity planning," Computers & Operations Research, vol. 40, no. 9, pp. 2198–2207, 2013.

[17] R. Burdett and E. Kozan, "A multi-criteria approach for hospital capacity analysis," European Journal of Operational Research, vol. 255, no. 2, pp.505–521, 2016.

[18] S. P. Siferd and W. Benton, "Workforce staffing and scheduling: Hospital nursing specific models," European Journal of Operational Research, vol. 60, no. 3, pp. 233–246, 1992.

[19] J. F. Bard and H. W. Purnomo, "Short-term nurse scheduling in response to daily fluctuations in supply and demand," Health Care Management Science, vol. 8, no. 4, pp. 315–324, 2005.

[20] K. Kim and S. Mehrotra, "A two-stage stochastic integer programming approach to integrated staffing and scheduling with application to nurse management," Operations Research, vol. 63, no. 6, pp. 1431–1451, 2015.

[21] N. Yankovic and L. V. Green, "Identifying good nursing levels: A queuing approach," Operations research, vol. 59, no. 4, pp. 942–955, 2011.

[22] P. Punnakitikashem, J. M. Rosenberber, and D. F. Buckley-Behan, "A stochastic programming approach for integrated nurse staffing and assignment," IISE Transactions, vol. 45, no. 10, pp. 1059–1076, 2013.

[23] Y. Li, Y. Zhang, N. Kong, and M. Lawley, "Capacity planning for long-term care networks," IIE Transactions, vol. 48, no. 12, pp. 1098–1111, 2016.

[24] Y. Zhang, M. L. Puterman, M. Nelson, and D. Atkins, "A simulation optimization approach to long-term care capacity planning," Operations research, vol. 60, no. 2, pp. 249–261, 2012.

[25] B. Efron and C. Morris, "Stein's estimation rule and its competitorsan empirical bayes approach," Journal of the American Statistical Association, vol. 68, no. 341, pp. 117–130, 1973.

[26] "CMS RUG-IV specifications and dll package," https://www.cms.gov/Medicare/Quality-Initiatives-Patient-Assessment-Instruments/NursingHomeQualityInits/NHQIMDS30TechnicalInformation.html, accessed 26 Jan 2021.

[27] K. van Eeden, D. Moeke, and R. Bekker, "Care on demand in nursing homes: a queueing theoretic approach," Health care management science, vol. 19, no. 3, pp. 227–240, 2016.

[28] I. S. Dhillon, S. Mallela, and R. Kumar, "A divisive information-theoretic feature clustering algorithm for text classification," Journal of machine learning research, vol. 3, no. Mar, pp. 1265–1287, 2003.

[29] J. Lin, "Divergence measures based on the shannon entropy," IEEE Transactions on Information theory, vol. 37, no. 1, pp. 145–151, 1991.

[30] M. A. Stephens, "Use of the kolmogorov–smirnov, cramer–von mises and related statistics without extensive tables," Journal of the Royal Statistical Society: Series B (Methodological), vol. 32, no. 1, pp. 115–122, 1970.

[31] R. Agrawal, R. Srikant et al., "Fast algorithms for mining association rules," in Proc. 20th int. conf. very large data bases, VLDB, vol. 1215, 1994, pp. 487–499.

[32] J. Gehrke, V. Ganti, R. Ramakrishnan, and W.-Y. Loh, "Boatoptimistic decision tree construction," in ACM SIGMOD Record, vol. 28, no. 2. ACM, 1999, pp. 169–180.

[33] A. J. Kleywegt, A. Shapiro, and T. Homem-de Mello, "The sample average approximation method for stochastic discrete optimization," SIAM Journal on Optimization, vol. 12, no. 2, pp. 479–502, 2002.

[34] D. Saliba, M. Jones, J. Streim, J. Ouslander, D. Berlowitz, and J. Buchanan, "Overview of significant changes in the minimum data set for nursing homes version 3.0," Journal of the American Medical Directors Association, vol. 13, no. 7, pp. 595–601, 2012.

[35] F. E. Harrell, R. M. Califf, D. B. Pryor, K. L. Lee, and R. A. Rosati, "Evaluating the yield of medical tests," Jama, vol. 247, no. 18, pp. 2543–2546, 1982.

[36] M. G. Akritas, "Bootstrapping the kaplanmeier estimator," Journal of the American Statistical Association, vol. 81, no. 396, pp. 1032–1038, 1986.